\documentclass[journal]{IEEEtran}
\usepackage{epsfig}
\usepackage{amsmath}
\usepackage{amsfonts}
\usepackage{amssymb}
\usepackage{multirow}
\usepackage{color}
\usepackage{cite}

\DeclareMathOperator*{\argmax}{arg\,max}

\newtheorem{proposition}{Proposition}

\newtheorem{rem}{Remark}

\begin{document}

\title{Coordinated Receding-Horizon Control of Battery Electric Vehicle Speed and Gearshift Using Relaxed Mixed Integer Nonlinear Programming}

\author{Nan Li, Kyoungseok Han$^{*}$, Ilya Kolmanovsky, and Anouck Girard 
\thanks{This research was supported by the National Science Foundation Award ECCS 1931738.}
\thanks{Nan Li, Ilya Kolmanovsky, and Anouck Girard are with the Department of Aerospace Engineering, University of Michigan, Ann Arbor, MI, 48109, USA (email: nanli@umich.edu, ilya@umich.edu, anouck@umich.edu)}
\thanks{Kyoungseok Han (Corresponding Author) is with the School of Mechanical Engineering, 
Kyungpook National University, Daegu 41566, Republic of Korea (email:kyoungsh@knu.ac.kr)}
}

% make the title area
\maketitle

\begin{abstract}
In this paper, we propose an approach to coordinated receding-horizon control of vehicle speed and transmission gearshift for automated battery electric vehicles (BEVs) to achieve improved energy efficiency. The introduction of multi-speed transmissions in BEVs creates an opportunity to manipulate the operating point of electric motors under given vehicle speed and acceleration command, thus providing the potential to further improve the energy efficiency. However, co-optimization of vehicle speed and transmission gearshift leads to a mixed integer nonlinear program (MINLP), solving which can be computationally very challenging. In this paper, we propose a novel continuous relaxation technique to treat such MINLPs that makes it possible to compute solutions with conventional nonlinear programming solvers. After analyzing its theoretical properties, we use it to solve the optimization problem involved in coordinated receding-horizon control of BEV speed and gearshift. Through simulation studies, we show that co-optimizing vehicle speed and transmission gearshift can achieve considerably greater energy efficiency than optimizing them sequentially, and the proposed relaxation technique can reduce the online computational cost to a level that is comparable to the time available for real-time implementation.
\end{abstract}

% Note that keywords are not normally used for peerreview papers.

\IEEEpeerreviewmaketitle

\section{Introduction}\label{sec:1}

\IEEEPARstart{W}{ith} the market penetration of Battery Electric Vehicles (BEVs) projected to increase, there has been a growing interest in approaches to  improving energy efficiency of BEVs. On the one hand, the increasing levels of connectivity and automation have opened up new opportunities to optimize the vehicle-level energy management \cite{vahidi2018energy,han2018safe,turri2016cooperative,sun2020optimal,ersal2020connected}. For instance, these technologies enable the vehicle to predict future road and traffic conditions, so the vehicle can take this information into account to plan its speed profile in a way that minimizes the energy consumption. On the other hand, further energy efficiency improvement may be achieved by adding components, such as multi-speed transmissions, to the existing BEV powertrain architecture. Traditionally, BEV powertrain only used a single reduction gear for forward driving \cite{dib2014optimal}. However, in recent years, multi-speed transmissions for electric vehicles have also been considered \cite{wager2014testing,ngo2012optimal,guo2016optimal,guo2017line}. With a multi-speed transmission, the operating point of the electric motor can be adjusted under given vehicle speed and acceleration command so that improved energy efficiency can be achieved at the powertrain level. 

In this paper, motivated by potential synergies in vehicle and powertrain-level energy management optimization, we consider co-optimization of vehicle speed and gearshift for BEVs equipped with multi-speed transmissions to maximize the energy efficiency.

The problem of energy management for BEVs has been studied by several researchers in the literature. Dynamic programming (DP) is a popular tool for determining the globally optimal trajectory offline when the entire trip is assumed to be known a priori \cite{ozatay2014cloud,mcdonough2014stochastic,zeng2018globally}. However, it has been revealed in \cite{asher2017prediction} that prediction errors of the future speed trajectory may have a significant impact on the energy consumption. Therefore, a practical solution is to repeatedly update the predictions based on the latest available traffic information to mitigate such errors, and also update the control correspondingly in real time. However, the intensive computations required by DP typically forbid its real-time implementation. Alternatively, Pontryagin's minimum principle (PMP) has been exploited for vehicle speed planning in \cite{saerens2013calculation,wan2016optimal}. A PMP-based energy management strategy for hybrid electric vehicles has also been proposed in \cite{ngo2012optimal}, which assumes a given speed profile and includes a gearshift strategy. Although PMP requires less computational effort than DP, the two-point boundary value problem associated with PMP conditions may still be difficult to handle numerically.

Using short-term preview information instead of prediction of the entire trip and optimizing vehicle speed based on a receding-horizon optimization/model predictive control framework has been considered in \cite{kamal2011ecological,kamal2012model,homchaudhuri2016fast,seok2018energy}. On the one hand, previewing road and traffic information over a short time horizon may be easier and also more accurate than prediction over a long horizon or over the entire trip \cite{lefevre2014comparison,hyeon2019short,liu2019vehicle}. On the other hand, the online optimization is reduced to a finite-dimensional mathematical programming problem, which is handled in real time. 

In this paper, we also consider a receding-horizon optimization strategy for improving BEV energy efficiency. In particular, we develop an approach to coordinated receding-horizon control of BEV speed and gearshift. Due to the discrete-valued gear ratios of a multi-speed transmission, the online problem representing the speed and gearshift co-optimization is a mixed integer nonlinear programming (MINLP) problem. To exactly solve such an MINLP problem is computationally very demanding \cite{belotti2013mixed}. For instance, conventional approaches based on exhaustive-search or branch-and-bound \cite{morrison2016branch} that either brute-force or systematically enumerate all solution candidates for the discrete variables have worst-case combinatorial complexity \cite{belotti2013mixed}. Therefore, optimizations of vehicle speed and transmission gearshift are often treated in a sequential/hierarchical manner in previous literature. In \cite{guo2016optimal}, the speed trajectory is optimized first based on a cost function that approximately represents the energy consumption but does not involve powertrain variables. The most energy efficient gear is then selected for the obtained speed trajectory. In our previous work \cite{han2020acc}, such a sequential procedure is augmented by an additional step, which refines the speed trajectory after the gearshift trajectory has been determined in the second step. Although these sequential optimization approaches appear to be effective in improving BEV energy efficiency based on simulation studies, optimality of the obtained trajectories is not guaranteed. 

In contrast, in this paper we treat the BEV speed and gearshift co-optimization problem in a receding-horizon control setting by first transforming it to an MINLP problem in a specific form, then proposing a novel continuous relaxation of this MINLP problem, and finally approximately solving the original MINLP problem through solving its corresponding continuous relaxation.

The contributions of this paper are as follows:

1) We propose a coordinated control strategy for BEV speed and gearshift based on receding-horizon optimization with specific cost and constraints.

2) We propose a novel continuous relaxation to the formulated MINLP that represents the speed and gearshift co-optimization problem. We first show that the relaxed problem is a nonlinear programming (NLP) problem with continuously differentiable cost and constraint functions and can be treated by off-the-shelf NLP solvers. We then show that the original MINLP problem and the relaxed NLP problem have no feasibility and optimality gaps at their global minimizers. Moreover, we also characterize the relationship between the local minimizers of the two problems and further address this through numerical studies.

3) We show based on a comprehensive set of simulation case studies that considerably greater energy efficiency can be achieved through co-optimization of vehicle speed and transmission gearshift than by optimizing them separately. We also show that the proposed relaxation technique can reduce the online computational cost to a level that is comparable to the time available for real-time implementation. Our approach thus opens up a possibility for real-time coordinated receding-horizon control of vehicle speed and transmission gearshift.

The rest of the paper is organized as follows: Section~\ref{sec:2} formulates the BEV speed and gearshift receding-horizon co-optimization problem and transforms the problem into an MINLP with a specific form. Section~\ref{sec:3} introduces the proposed continuous relaxation technique to the transformed MINLP problem and discusses its theoretical properties. Section~\ref{sec:4} describes several other energy efficiency optimization approaches for BEV speed and gearshift control to compare versus the proposed receding-horizon co-optimization strategy. A comprehensive set of simulation results is presented in Section~\ref{sec:5}. Finally, Section~\ref{sec:6} concludes the paper.

\section{Battery Electric Vehicle Speed and Gearshift Co-Optimization}\label{sec:2}

In this section, we formulate the BEV speed and gearshift co-optimization problem for improved energy efficiency. We first introduce the models for vehicle longitudinal motion, battery state-of-charge ($SOC$) and transmission gear dynamics, and then define the co-optimization problem.

\subsection{Vehicle and Battery Models}\label{sec:21}

The longitudinal motion of the vehicle is modeled as follows:
\begin{align} 
    &\dot{s}=v, \label{equ:system_model} \\ 
    &\dot{v}=\frac{T_{\text{w}}}{r_{\text{w}} m_{\text{eff}}}-\frac{1}{2m}\rho A_{\text{f}} C_{\text{d}} v^2 - g (\sin\theta + \mu \cos\theta), \label{equ:system_model2}
\end{align}
where $s$ is the vehicle travel distance, $v$ is the vehicle speed, 
$m$ is the vehicle mass, $m_{\text{eff}}$ is the vehicle effective mass accounting for both static mass and rotational inertia effects, $T_{\text{w}}$ is the wheel torque, $r_{\text{w}}$ is the tire radius, $\rho$ is the air density, $A_{\text{f}}$ is the vehicle frontal area, $C_{\text{d}}$ is the aerodynamic drag coefficient, $g$ is the gravitational constant, $\theta$ is the road inclination, and $\mu$ is the rolling resistance coefficient.

The wheel torque $T_{\text{w}}$ is determined by the motor torque $T_{\text{m}}$, friction brake torque $T_{\text{b}}$, reduction gear ratio $i_{\text{g}}$, and final drive ratio $i_0$ as follows:
\begin{equation}\label{equ:wheel_torque}
    T_{\text{w}} = T_{\text{m}}\, i_{\text{g}}\, i_0 - T_{\text{b}}.
\end{equation}
Ideally, friction brakes are used only when the maximum torque that can be provided by the motor is not sufficient to achieve the required braking. In this paper, we assume that such cases are not occurring, i.e., $T_{\text{b}} = 0$.

The evolution of battery $SOC$ is modeled as \cite{guo2017line}:
\begin{equation}\label{equ:SoC}
    \dot{SOC} = -\frac{I_{\text{b}}}{C}=-\frac{V_{\text{oc}}-\sqrt{V_{\text{oc}}^2-4R_{\text{b}} P_{\text{b}}}}{2\,C R_{\text{b}}},
\end{equation}
where $I_{\text{b}}$ is the battery current, $C$ is the battery capacity, $V_{\text{oc}} = V_{\text{oc}}(SOC)$ is the open-circuit voltage in series with the battery resistance $R_{\text{b}} = R_{\text{b}}(SOC)$, both of which depend on battery $SOC$, and $P_{\text{b}}$ is the battery power determined as:
\begin{equation}\label{equ:Pb}
    P_{\text{b}} = \begin{cases} \frac{P_{\text{m}}}{\eta_{\text{b}}^+} = \frac{T_{\text{m}} w_{\text{m}}}{\eta_{b}^+ \eta_{\text{m}}} & \text{when } T_{\text{m}} \geq 0, \\[6pt]
    \frac{P_{\text{m}}}{\eta_{\text{b}}^-} = \frac{T_{\text{m}} w_{\text{m}}}{\eta_{b}^- \eta_{\text{m}}} & \text{when } T_{\text{m}} < 0, \end{cases}
\end{equation}
where $\eta_{\text{b}}^+ \in (0,1)$ is the battery-depletion efficiency, $\eta_{\text{b}}^- > 1$ is the battery-recharge efficiency,  $w_{\text{m}}$ is the motor speed determined as:
\begin{equation}\label{equ:motor_speed}
    w_{\text{m}} = \frac{v}{r_{\text{w}}}\, i_{\text{g}}\, i_0,
\end{equation}
and $\eta_{\text{m}}$ is the motor efficiency depending on the motor operating point, i.e., $\eta_{\text{m}} = \eta_{\text{m}} (T_{\text{m}},w_{\text{m}})$. The maps $V_{\text{oc}}(SOC)$, $R_{\text{b}}(SOC)$ and $\eta_{\text{m}} (T_{\text{m}},w_{\text{m}})$ are typically estimated based on experimental data and provided as lookup tables.

We discretize the continuous-time models \eqref{equ:system_model}-\eqref{equ:motor_speed} using the forward Euler method with a sampling period $\Delta t$, and combine them into a discrete-time model in the form of
\begin{equation}\label{equ:system_model3}
    \xi_{t+1} = \Phi(\xi_t,\gamma_t),
\end{equation}
where $\xi = [s,v,SOC]^{\top}$ is the state vector, $\gamma = [T_{\text{m}},i_{\text{g}}]^{\top}$ is the input vector, the subscript $t$ represents the discrete time instant, and the function $\Phi$ is determined by \eqref{equ:system_model}-\eqref{equ:motor_speed} and the sampling period $\Delta t$.

\subsection{Transmission Model}\label{sec:22}

For a BEV with a multi-speed transmission, the gear ratio $i_{\text{g}}$ can take a finite number of different values. In particular, for a given transmission model with $\eta_{\max}$ gears,  $i_{\text{g}}$ is determined by the gear position $\eta_{\text{g}}$, i.e., $i_{\text{g}} = i_{\text{g}}(\eta_{\text{g}})$, with $\eta_{\text{g}} \in \{1,\cdots,\eta_{\max}\}$. To avoid gear skipping, we model gear changes as \cite{guo2017line}:
\begin{equation}\label{equ:gearshift}
  \eta_{\text{g},t+1} = \eta_{\text{g},t} + \zeta_t,
\end{equation}
where $\zeta_t \in \{-1,0,1\}$ is the gearshift signal, with $-1$ and $1$ representing, respectively, the down- and up- shift signals, and $0$ representing maintaining the current gear position. In this paper, a three-speed transmission is assumed to be employed, i.e., $\eta_{\max} = 3$.

\subsection{Speed and Gearshift Receding-Horizon Co-Optimization}\label{sec:23}

We pursue co-optimization of vehicle speed and transmission gearshift to maximize energy efficiency. Considering an automated vehicle control system, the reference vehicle speed $v_{\text{r}}$ and the corresponding reference travel distance $s_{\text{r}}$ are typically available over a short time horizon, however, deviations from these reference trajectories are permissible within prescribed bounds. Such a reference speed and distance preview can be informed by short-term prediction of the speed(s) of the vehicle(s) driving in front \cite{lefevre2014comparison,hyeon2019short,liu2019vehicle} or by an automated vehicle planning module \cite{schwarting2018planning}. In this paper, we assume $v_{\text{r}}$ corresponds to the predicted speed of the vehicle driving immediately in front of and being followed by the ego BEV.

The speed and gearshift co-optimization is achieved by solving the following minimization problem repeatedly in a receding-horizon manner:
\begin{subequations}\label{equ:hyb_opt_prob}
\begin{align} 
    \min \quad & J_{t} = -SOC_{N|t} + \sum_{k=0}^{N-1} \big( w_1\,(v_{k+1|t}-v_{\text{r},k+1|t})^2 \nonumber \\[-1pt]
    &\quad\quad + w_2\,(T_{\text{w},k|t}-T_{\text{w},k-1|t})^2 \big), \label{equ:hyb_opt_prob_cost} \\[3pt]
            \text{s.t.}\quad &\quad\quad \xi_{k+1|t} = \Phi(\xi_{k|t},\gamma_{k|t}), \label{equ:hyb_opt_prob_c1} \\[2pt] 
                & \tau_{\min} (v_{k+1|t}+\delta_1) \le s_{\text{r},k+1|t} - s_{k+1|t} \nonumber \\
                &\quad\quad\quad\quad\quad\quad \le \tau_{\max} (v_{k+1|t}+\delta_1), \label{equ:hyb_opt_prob_c3} \\[2pt]
                &\quad |v_{k+1|t}-v_{\text{r},k+1|t}| \leq \max(\varepsilon\, v_{\text{r},k+1|t}, \delta_2), \label{equ:hyb_opt_prob_c4}  \\[2pt]
                & T_{\min} (w_{\text{m},k|t}) \leq T_{\text{m},k|t} \leq T_{\max}(w_{\text{m},k|t}), \label{equ:hyb_opt_prob_c5} \\[2pt]
                &\quad\quad i_{\text{g},k|t} = i_{\text{g}}(\eta_{\text{g},k|t}), \label{equ:hyb_opt_prob_g1} \\[2pt]
                &\quad\quad \eta_{\text{g},k+1|t} = \eta_{\text{g},k|t} + \zeta_{k|t} \label{equ:hyb_opt_prob_g2} \\[2pt]
                &\quad\quad \eta_{\text{g},k+1|t} \in \{1,\cdots,\eta_{\max}\}, \label{equ:hyb_opt_prob_g3} \\[2pt]
                &\quad\quad\quad \zeta_{k|t} \in \{-1,0,1\}, \label{equ:hyb_opt_prob_g4} \\[2pt]
                &\quad\quad\quad\quad\quad\quad\quad  \forall\, k =0, \cdots, N-1, \nonumber \\[-8pt]
                &\quad\quad\quad \sum_{k=0}^{N-1} |\zeta_{k|t}| \le \zeta_{\max}, \label{equ:hyb_opt_prob_g5}
\end{align} 
\end{subequations}
with respect to the decision variables $u_{k|t} = [T_{\text{m},k|t},\zeta_{k|t}]^{\top}$, $k = 0,\cdots,N-1$, where the notation $(\cdot)_{k|t}$ designates a predicted value of the variable $(\cdot)_{t+k}$ with the prediction made at the current time instant $t$. 

The motor torque $T_{\text{m}}$ and the gearshift signal $\zeta$ are chosen as the decision variables because the values of all other variables, including the vehicle speed $v$ and the battery $SOC$, can be uniquely determined by $T_{\text{m}}$ and $\zeta$ based on the models \eqref{equ:system_model3} and \eqref{equ:gearshift}, which are treated as equality constraints in \eqref{equ:hyb_opt_prob_c1} and \eqref{equ:hyb_opt_prob_g2}. The term $-SOC_{N|t}$ in the cost function \eqref{equ:hyb_opt_prob_cost} is for minimizing energy consumption. The terms $(v_{k+1|t}-v_{\text{r},k+1|t})^2$ and $(T_{\text{w},k|t}-T_{\text{w},k-1|t})^2 = (T_{\text{m},k|t}\, i_{\text{g},k|t} - T_{\text{m},k-1|t}\, i_{\text{g},k-1|t})^2 i_0^2$ in \eqref{equ:hyb_opt_prob_cost} are for penalizing deviations of the actual speeds from the reference speeds and for penalizing changes in wheel torques, respectively, to improve safety and comfort (by reducing jerk). The constraint \eqref{equ:hyb_opt_prob_c3} represents the requirement of keeping the ego BEV's time-headway to its preceding vehicle within the range $[\tau_{\min},\tau_{\max}]$ to avoid rear-end collisions and cut-ins by other vehicles, where the predicted travel distances of the preceding vehicle $s_{\text{r},k+1|t}$ are determined according to the dynamic equation $s_{\text{r},k+1|t} = s_{\text{r},k|t} + v_{\text{r},k|t}\, \Delta t$ based on the current distance $s_{\text{r},0|t} = s_{\text{r},t}$ and the predicted speeds $v_{\text{r},k|t}$. The constraint \eqref{equ:hyb_opt_prob_c4} represents prescribed bounds on the maximum deviations of the actual speeds from the reference speeds. The constraint \eqref{equ:hyb_opt_prob_c5} represents the range of torques, $[T_{\min},T_{\max}]$, that can be provided by the motor at the speed $w_{\text{m},k|t}$. The constraints \eqref{equ:hyb_opt_prob_g1}-\eqref{equ:hyb_opt_prob_g4} correspond to the transmission model introduced in Section~\ref{sec:22}. And finally, the constraint \eqref{equ:hyb_opt_prob_g5} requires the number of gearshifts over the planning horizon to be upper bounded by $\zeta_{\max}$, to avoid overly frequent gearshifts.

At every discrete time instant $t$, after solving the optimization problem \eqref{equ:hyb_opt_prob}, the ego BEV applies the obtained $T_{\text{m},0|t}$ and $\zeta_{0|t}$ over one sampling period $\Delta t$ to update its states, then repeats this procedure at the next time instant $t+1$.

Due to the fact that $T_{\text{m},k|t}$ takes continuous values and $\zeta_{k|t}$ takes values in the discrete set $\{-1,0,1\}$, the problem \eqref{equ:hyb_opt_prob} is a mixed integer problem with many constraints. Note that in practice the number of gearshifts over a short time period (e.g., $5 \sim 8$ seconds) is typically small \cite{ngo2012optimal,guo2016optimal,guo2017line}. This means that the maximum number of gearshifts $\zeta_{\max}$ in \eqref{equ:hyb_opt_prob_g5} can be chosen as a small positive integer. For such a case, we introduce a transformation of \eqref{equ:hyb_opt_prob} that has fewer decision variables and constraints in what follows.

\subsection{Problem Transformation}\label{sec:24}

For a given gear position at the current time instant, $\eta_{\text{g},0|t} = \eta_{\text{g},t} \in \{1,\cdots,\eta_{\max}\}$, there are a finite number of distinct gear position sequences, $\pi_t = \{\eta_{\text{g},0|t}, \eta_{\text{g},1|t},\cdots,\eta_{\text{g},N|t}\}$, that satisfy both the gear dynamics \eqref{equ:hyb_opt_prob_g2}-\eqref{equ:hyb_opt_prob_g4} and the bound \eqref{equ:hyb_opt_prob_g5} on the number of gearshifts. We denote the set of all such sequences as $\Pi(\eta_{\text{g},t})$, called the set of admissible gear position sequences. We have that 1) $\pi_t$ takes values in $\Pi(\eta_{\text{g},t})$, and 2) $\Pi(\eta_{\text{g},t}) \in \big\{\Pi(1),\cdots, \Pi(\eta_{\max})\big\}$, where the sets $\Pi(1),\cdots, \Pi(\eta_{\max})$ can be constructed offline and stored for online use. 

Then, we can transform \eqref{equ:hyb_opt_prob} into the following problem:
\begin{subequations}\label{equ:hyb_opt_prob_tf}
\begin{align} 
    \min &\quad\quad  \eqref{equ:hyb_opt_prob_cost} \\[3pt]
            \text{s.t.} &\quad\quad \eqref{equ:hyb_opt_prob_c1}-\eqref{equ:hyb_opt_prob_g1} \\[2pt]
                &\,\, \{\eta_{\text{g},0|t}, \eta_{\text{g},1|t},\cdots,\eta_{\text{g},N|t}\} = \pi_t \in \Pi(\eta_{\text{g},t}), \label{equ:hyb_opt_prob_g_tf}
\end{align} 
\end{subequations}
with respect to the decision variables $T_{\text{m},k|t}$, $k = 0,\cdots,N-1$, and $\pi_t$. 

Moreover, after indexing the admissible gear position sequences $\pi_t$ in $\Pi(\eta_{\text{g},t})$ by natural numbers $1,2,\cdots$, we can write the problem \eqref{equ:hyb_opt_prob_tf} into the abstract form \eqref{equ:MINLP}. We remark that to achieve such an abstraction, the intermediate variables $\xi_{k|t}$ and $i_{\text{g},k|t}$ governed by the equality constraints \eqref{equ:hyb_opt_prob_c1} and \eqref{equ:hyb_opt_prob_g1} need to be considered as deterministic functions of the decision variables $T_{\text{m},k|t}$ and $\pi_t$. This way, not only the equality constraints \eqref{equ:hyb_opt_prob_c1} and \eqref{equ:hyb_opt_prob_g1} can be dropped from the problem definition, but also the inequality constraints \eqref{equ:hyb_opt_prob_c3}-\eqref{equ:hyb_opt_prob_c5} can be treated as conditions directly constraining the decision variables $T_{\text{m},k|t}$ and $\pi_t$ (i.e., by substituting the expressions of $\xi_{k|t}$ and $i_{\text{g},k|t}$ as functions of $T_{\text{m},k|t}$ and $\pi_t$ into them). Furthermore, the constraint \eqref{equ:hyb_opt_prob_c4} needs to be expressed as two inequalities so that each of them involves a continuously differentiable function. 

In the next section, we deal with the problem \eqref{equ:hyb_opt_prob_tf} by looking at its abstract, condensed form \eqref{equ:MINLP}.

\section{A Continuous Relaxation to a Mixed Integer Optimization Problem}\label{sec:3}

In Section~\ref{sec:24}, we transformed the BEV speed and gearshift co-optimization problem into the following form, which is an MINLP:
\begin{subequations}\label{equ:MINLP}
\begin{align}
    \min_{u,v} \quad & f(u,v), \label{equ:MINLP_cost} \\
    \quad \begin{split}
    \text{s.t.}\quad & u \in U(v) = \{u \in \mathbb{R}^{n_u} : g(u,v) \le \boldsymbol{0}_m \}, \\[2pt]
    & v \in V = \{1,\cdots,n_v\}, \end{split} \label{equ:MINLP_constraint}
\end{align}
\end{subequations}
where the cost function $f(u,v): \mathbb{R}^{n_u} \times \mathbb{N} \to \mathbb{R}$ is assumed to be continuously differentiable in $u$. The continuous variable $u$ takes values in a set $U(v)$, which depends on $v$ and is characterized by the inequalities $g(u,v) \leq \boldsymbol{0}_m$ with $g: \mathbb{R}^{n_u} \times \mathbb{N} \to \mathbb{R}^m$ being continuously differentiable in $u$. The integer variable $v$ takes values in a finite set $V = \{1,\cdots,n_v\} \subset \mathbb{N}$.

In general, MINLP problems are difficult to solve exactly. Continuous relaxation techniques may be exploited to obtain approximate solutions \cite{axehill2010convex}. They typically transform the original MINLP problem into a nonlinear programming problem with purely continuous variables (NLP). For instance, for some problems, the integrality constraint $v \in V = \{1,\cdots,n_v\}$ may be replaced with $v \in \bar{V} = [1,n_v]$. Then, one can use off-the-shelf NLP solvers, such as the interior-point method \cite{potra2000interior} and the sequential quadratic programming (SQP) method \cite{gill2012sequential}, to compute solutions to the transformed problem.

However, in our BEV speed and gearshift co-optimization problem, neither a gear ratio $i_{\text{g}}(\eta_{\text{g}})$ with a non-integer gear position $\eta_{\text{g}}$ (see \eqref{equ:hyb_opt_prob_g1}) nor a gear position sequence $\pi_t \in \Pi(\eta_{\text{g},t})$ with a non-integer index (see \eqref{equ:hyb_opt_prob_g_tf}) are defined. This means a model that represents the powertrain response with such non-integer settings is unavailable. In this case, the above relaxation where $v \in V = \{1,\cdots,n_v\}$ is replaced with $v \in \bar{V} = [1,n_v]$ is not applicable to our problem. Therefore, in what follows we introduce another continuous relaxation and also discuss its theoretical properties.

Firstly, it is easy to see that the problem \eqref{equ:MINLP} is equivalent to the following MINLP problem:
\begin{subequations}\label{equ:equivalent_MINLP}
\begin{align}
    \min_{u,p} \quad & p^{\top} {\bf f}(u), \\
    \quad \begin{split}  \text{s.t.} \quad & {\bf g}(u)\, p \le \boldsymbol{0}_{m n_v}, \\[2pt]
    & p \in \Omega, \end{split} \label{equ:equivalent_MINLP_c} 
\end{align}
\end{subequations}
where
\begin{equation}
{\bf f}(u) = \begin{bmatrix} f(u,1) \\ \vdots \\ f(u,n_v) \end{bmatrix},\quad {\bf g}(u) = \begin{bmatrix} g(u,1) & & \\ & \ddots & \\ & & g(u,n_v) \end{bmatrix},
\end{equation}
and
\begin{equation}\label{equ:Omega}
\Omega = \big\{p \in \{0,1\}^{n_v}  \,|\, p^{\top} \boldsymbol{1}_{n_v} = 1 \big\}.
\end{equation}
The set $\Omega$ defined in \eqref{equ:Omega} represents the set of vertices of an $(n_v-1)$-dimensional standard simplex, and the constraint $p \in \Omega$ ensures that the vector $p$ has precisely one entry to be $1$ and all others to be $0$. In particular, the index of the entry $1$ corresponds to the value of $v$ in \eqref{equ:MINLP}.

We consider the following continuous relaxation to \eqref{equ:equivalent_MINLP}:
\begin{subequations}\label{equ:relax_MINLP}
\begin{align}
    \min_{u,p} \quad & p^{\top} {\bf f}(u), \\
     \quad \begin{split} \text{s.t.} \quad &  {\bf g}(u)\, p \le \boldsymbol{0}_{m n_v}, \\[2pt] 
    & p \in \bar{\Omega}, \end{split} \label{equ:relax_MINLP_c} 
\end{align}
\end{subequations}
where
\begin{equation}\label{equ:Omega_bar}
\bar{\Omega} = \big\{p \in [0,1]^{n_v}  \,|\, p^{\top} \boldsymbol{1}_{n_v} = 1 \big\}.
\end{equation}
It can be seen that \eqref{equ:relax_MINLP} is transformed from \eqref{equ:equivalent_MINLP} by replacing the vertex set $\Omega$ with its convex hull $\bar{\Omega}$. We now discuss several theoretical properties of the relaxed problem \eqref{equ:relax_MINLP}.

\begin{proposition}
The cost and constraint functions of \eqref{equ:relax_MINLP} are continuously differentiable in the decision variables $(u,p)$.
\end{proposition}

{\it Proof:} This follows from the expressions of the cost and constraints in \eqref{equ:relax_MINLP} and \eqref{equ:Omega_bar}, and our assumptions that $f$ and $g$ are continuously differentiable in $u$ made when problem \eqref{equ:MINLP} is defined. $\blacksquare$

The significance of Proposition~1 is that the continuous differentiability of the cost and constraint functions enables us to use derivative information to characterize minimizers of \eqref{equ:relax_MINLP}, e.g., through the Karush-Kuhn-Tucker conditions. This also implies that many off-the-shelf NLP solvers can be used to solve \eqref{equ:relax_MINLP}.

We are interested in characterizing the feasibility and optimality gaps between the original MINLP problem \eqref{equ:MINLP} and the relaxed problem \eqref{equ:relax_MINLP}. The following two propositions are dedicated to such properties.

\begin{proposition}
Suppose $(\bar{u},\bar{p})$ is a global minimizer of \eqref{equ:relax_MINLP}. Let $\hat{p} \in \Omega$ be such that $\hat{p}_i = 1$ for some $i$ satisfying $\bar{p}_i>0$. Then, $(\bar{u},\hat{p})$ is necessarily a global minimizer of \eqref{equ:equivalent_MINLP}. In turn, $(\bar{u},i)$ is a global minimizer of \eqref{equ:MINLP}. Moreover, we have $\bar{p}^{\top} {\bf f}(\bar{u}) = \hat{p}^{\top} {\bf f}(\bar{u}) = f(\bar{u},i)$.
\end{proposition}

{\it Proof:} Let $V' = \{j \in V: g(\bar{u},j) \le \boldsymbol{0}_{m}\}$. Since $(\bar{u},\bar{p})$ is feasible for \eqref{equ:relax_MINLP}, i.e., $g(\bar{u},k) \le \boldsymbol{0}_{m}$ for all $k \in \{j \in V: \bar{p}_j>0\}$, the set $V'$ is non-empty. Let us rename the integers in $V'$ as $1,2,\cdots,m_v,m_v+1$ with $0 \le m_v \le n_v-1$.

Let $\Sigma = \big\{q \in [0,1]^{m_v}  \,|\, q^{\top} \boldsymbol{1}_{m_v} \le 1 \big\}$ and $\bar{\Omega}' = \Bigg\{\begin{bmatrix} q \\ 1-q^{\top} \boldsymbol{1}_{m_v} \\ \boldsymbol{0}_{n_v-m_v-1} \end{bmatrix} \Bigg|\, q \in \Sigma \Bigg\} \subseteq \bar{\Omega}$. By construction, $(\bar{u},p')$ is feasible for \eqref{equ:relax_MINLP} if and only if $p' \in \bar{\Omega}'$. In particular, $\bar{p} \in \bar{\Omega}'$. Let us write $\bar{p}$ as $\begin{bmatrix} \bar{q} \\ 1-\bar{q}^{\top} \boldsymbol{1}_{m_v} \\ \boldsymbol{0}_{n_v-m_v-1} \end{bmatrix}$ with $\bar{q} \in \Sigma$.

Let us now consider $h: \mathbb{R}^{m_v} \to \mathbb{R}$ defined by 
\begin{equation}\label{equ:g}
h(q) = \sum_{k = 1}^{m_v} q_k f(\bar{u},k) + (1-q^{\top} \boldsymbol{1}_{m_v}) f(\bar{u},m_v+1),
\end{equation}
which is a linear function on $\mathbb{R}^{m_v}$. Since $(\bar{u},\bar{p})$ is a global minimizer of \eqref{equ:relax_MINLP}, $\bar{q}$ is a global minimizer of $h$ on $\Sigma$, which is a compact subset of $\mathbb{R}^{m_v}$.

Since $h$ is linear, either $\bar{q}$ locates on the boundary of $\Sigma$ or $h$ is constant on $\Sigma$.\footnote{This follows from the fact that linear functions are harmonic and the maximum principle for harmonic functions \cite{axler2013harmonic}.} For the former case, either $\bar{q}_j \in \{0,1\}$ for some $j \in \{1,\cdots,m_v\}$ or $\bar{q}^{\top} \boldsymbol{1}_{m_v} = 1$. 

If $\bar{q}_j = 1$, then $\bar{p}$ satisfies $\bar{p}_j = 1$ and $\bar{p}_k = 0$ for all $k \neq j$. In this case, the $\hat{p}$ defined in the proposition statement is identical to $\bar{p}$, and $(\bar{u},\hat{p}) = (\bar{u},\bar{p})$ is feasible for both the MINLP problem \eqref{equ:equivalent_MINLP} and the relaxed problem \eqref{equ:relax_MINLP}. Moreover, we have $\hat{p}^{\top} {\bf f}(\bar{u}) = \bar{p}^{\top} {\bf f}(\bar{u})$.

If $\bar{q}_j = 0$ which implies $\bar{p}_j = 0$ or $\bar{q}^{\top} \boldsymbol{1}_{m_v} = 1$ which implies $\bar{p}_{m_v+1} = 0$, then we exclude $j$ or $m_v+1$ from $V'$, rename the remaining integers in $V'$ as $1,2,\cdots,m_v',m_v'+1$ with $m_v' = m_v-1$, and repeat the above arguments. By iterating this procedure, we will eventually fall into the case where $h$ is constant on $\Sigma$. Specifically, we will end up with a maximum set of integers $V'' = \{1,\cdots,m_v'',m_v''+1\}$ and its corresponding $\bar{\Omega}'' = \Bigg\{\begin{bmatrix} q \\ 1-q^{\top} \boldsymbol{1}_{m_v''} \\ \boldsymbol{0}_{n_v-m_v''-1} \end{bmatrix} \Bigg|\, q \in [0,1]^{m_v''},\, q^{\top} \boldsymbol{1}_{m_v''} \le 1 \Bigg\}$ such that 1) $f(\bar{u},j) = f(\bar{u},k)$ for $j,k \in V''$, and 2) $f(\bar{u},j) \le f(\bar{u},k)$ for $j \in V''$ and $k \in V' \setminus V''$. In particular, $\bar{p} \in \bar{\Omega}''$.

The above 1) and 2) show that when the continuous variable $u$ is fixed at the given value $\bar{u}$, every $\hat{p} \in \Omega$ such that $\hat{p}_i = 1$ only if $\bar{p}_i>0$ is a global minimizer of the induced integer program. Note that by construction, such $\hat{p}$'s must belong to $\bar{\Omega}''$. Moreover, we have that $\bar{p}^{\top} {\bf f}(\bar{u})$, as a convex combination of identical $f(\bar{u},j)$'s with $j \in V''$, satisfies $\bar{p}^{\top} {\bf f}(\bar{u}) = \hat{p}^{\top} {\bf f}(\bar{u})$ for every such $\hat{p}$.

Since the admissible set defined by \eqref{equ:equivalent_MINLP_c}, $\Xi$, is a subset of that defined by \eqref{equ:relax_MINLP_c}, $\bar{\Xi}$, and $\hat{p}^{\top} {\bf f}(\bar{u}) = \bar{p}^{\top} {\bf f}(\bar{u}) \le p^{\top} {\bf f}(u)$ for all $(u,p) \in \bar{\Xi}$, it must hold that $\hat{p}^{\top} {\bf f}(\bar{u}) \le p^{\top} {\bf f}(u)$ for all $(u,p) \in \Xi \subset \bar{\Xi}$, i.e., $(\bar{u},\hat{p})$ is a global minimizer of \eqref{equ:equivalent_MINLP}.

The remaining part of Proposition~2 follows from the equivalence of \eqref{equ:MINLP} and \eqref{equ:equivalent_MINLP}.
$\blacksquare$

Proposition~2 says that the original MINLP problem \eqref{equ:MINLP} and the relaxed problem \eqref{equ:relax_MINLP} have no optimality gap at their global minimizers. Furthermore, if a global minimizer  $(\bar{u},\bar{p})$ to the relaxed problem \eqref{equ:relax_MINLP} has been found, it is straightforward to derive a global minimizer $(\bar{u},i)$ to the original MINLP problem \eqref{equ:MINLP}. In practice, it may not be easy to identify a global minimizer to a non-convex NLP problem such as \eqref{equ:relax_MINLP}. Instead, typical NLP solvers, especially the ones exploiting derivative information, compute only local minimizers. Therefore, we are also interested in estimating the gap between \eqref{equ:MINLP} and \eqref{equ:relax_MINLP} at their local minimizers. The following proposition presents such a result.

\begin{proposition}
Suppose $(\bar{u},\bar{p})$ is a local minimizer of \eqref{equ:relax_MINLP}. Let $\hat{p} \in \Omega$ be such that $\hat{p}_i = 1$ for some $i$ satisfying $\bar{p}_i>0$. Then, (i) $(\bar{u},\hat{p})$ and $(\bar{u},i)$ are guaranteed to be feasible points of \eqref{equ:equivalent_MINLP} and \eqref{equ:MINLP}, respectively, and, $\bar{p}^{\top} {\bf f}(\bar{u}) = \hat{p}^{\top} {\bf f}(\bar{u}) = f(\bar{u},i)$. (ii) If $\bar{p} \in \Omega$, then we have $\hat{p} = \bar{p}$, and $(\bar{u},\hat{p})$ and $(\bar{u},i)$ are guaranteed to be local minimizers of \eqref{equ:equivalent_MINLP} and \eqref{equ:MINLP}, respectively.
\end{proposition}

{\it Proof:} The proof for part (i) follows the same steps as those in the proof of Proposition~2 except for the second last paragraph. If $\bar{p}$ is itself an integer vector (i.e., $\bar{p} \in \Omega$), then $\hat{p}$ must be identical to $\bar{p}$ by its definition. Moreover, the fact that $(\bar{u},\bar{p})$ is a local minimizer of \eqref{equ:relax_MINLP} ensures $\bar{u}$ to be a local minimizer of the NLP problem with respect to $u$ that is induced from \eqref{equ:relax_MINLP} by fixing $p = \bar{p}$. The part (ii) of Proposition~3 thus follows.
$\blacksquare$

The significance of Proposition~3 is that it implies one can find a feasible solution $(\bar{u},i)$ to the MINLP problem \eqref{equ:MINLP} by rounding a solution $(\bar{u},\bar{p})$ to the NLP problem \eqref{equ:relax_MINLP}, which can be relatively easily solved for (e.g., using off-the-shelf NLP solvers). Moreover, the quality of such a rounded solution can be monitored via the cost value $\bar{p}^{\top} {\bf f}(\bar{u})$ of the solution to the NLP problem. In addition, the following two remarks discuss the optimality of the rounded solution $(\bar{u},i)$.

\begin{rem}
In general, the rounded solutions $(\bar{u},\hat{p})$ and $(\bar{u},i)$ are not necessarily local minimizers of \eqref{equ:equivalent_MINLP} and \eqref{equ:MINLP}. This can be seen from the following example:
\begin{align}\label{equ:Ex_1}
    \min_{u,p_1,p_2} & \quad (p_1 + p_2) u, \\
    \,\,\,\, \begin{split}\text{s.t.}\quad & \begin{bmatrix} -p_1 u \\ -p_2 (u+1) \end{bmatrix} \le \boldsymbol{0}_2, \\[2pt] 
    & p_1, p_2 \in \{0,1\}, \quad p_1 + p_2 = 1, \end{split} \nonumber
\end{align}
which has the following continuous relaxation,
\begin{align}\label{equ:Ex_1_relax}
    \min_{u,p_1,p_2} & \quad (p_1 + p_2) u, \\
    \,\,\,\, \begin{split}\text{s.t.}\quad & \begin{bmatrix} -p_1 u \\ -p_2 (u+1) \end{bmatrix} \le \boldsymbol{0}_2, \\[2pt] 
    & p_1, p_2 \in [0,1], \quad p_1 + p_2 = 1. \end{split} \nonumber
\end{align}
It can be easily checked that $(\bar{u},\bar{p}_1,\bar{p}_2) = (0,0.5,0.5)$ is a local minimizer of \eqref{equ:Ex_1_relax}, but one of its rounded solutions, $(0,0,1)$, is not a local minimizer of \eqref{equ:Ex_1}. Indeed, $(-1,0,1)$ is the global minimizer of both \eqref{equ:Ex_1} and \eqref{equ:Ex_1_relax}, consistent with our theoretical result of Proposition~2.
\end{rem}

\begin{rem}
In the proof of Proposition~2, we have also shown that for $(\bar{u},\bar{p})$ with a non-integer $\bar{p}$ to be a (global or local) minimizer of \eqref{equ:relax_MINLP}, there must exist distinct vertices $i,j \in V$ of $\Omega$ such that $f(\bar{u},i) = f(\bar{u},j)$. This can rarely be encountered in many practical problems. For instance, we will show in Section~\ref{sec:5} that the numerical solutions to the relaxed version of our speed and gearshift receding-horizon co-optimization problem \eqref{equ:hyb_opt_prob_tf} have integer $\bar{p}$'s at the majority of the time instants. Proposition~3 ensures that these solutions with integer $\bar{p}$'s are guaranteed to be local minimizers of the original MINLP problem \eqref{equ:hyb_opt_prob_tf}.
\end{rem}

On the basis of Propositions~1--3, we approximately solve the BEV speed and gearshift co-optimization problem \eqref{equ:hyb_opt_prob_tf} through solving its continuous relaxation with the form \eqref{equ:equivalent_MINLP}. In particular, we determine the value of the integer variable according to $\bar{v} = \argmax_{i \in V} \bar{p}_i$.

\section{Alternative Energy Efficiency Optimization Approaches for Comparison}\label{sec:4}

To evaluate performance of the proposed vehicle speed and transmission gearshift coordinated receding-horizon control strategy for improving BEV energy efficiency, in this section we describe several other energy efficiency optimization approaches for comparison.

Firstly, we consider the globally optimal solution for vehicle speed and transmission gearshift trajectories in terms of minimizing battery \textit{SOC} consumption. It is computed as the solution to the following optimization problem:
\begin{subequations}\label{equ:DP}
\begin{align}
    \min \quad & J^{\text{dp}} = -SOC_{t_{\text{f}}}, \\[2pt]
    \text{s.t.} \quad & \eqref{equ:hyb_opt_prob_c1}-\eqref{equ:hyb_opt_prob_g4},
\end{align}
\end{subequations}
with respect to $u_{t} = [T_{\text{m},t},\zeta_{t}]^{\top}$, $t = 0,\cdots,t_{\text{f}}-1$, where $t_{\text{f}}$ corresponds to the entire duration of the trip. 

We remark that to compute such a globally optimal solution based on \eqref{equ:DP}, the reference speed $v_{\text{r}}$ over the entire trip must be known a priori. This may not be as practical as the assumption of being able to predict $v_{\text{r}}$ for a short time horizon in the proposed receding-horizon optimization strategy \eqref{equ:hyb_opt_prob}. In addition, the solution to \eqref{equ:DP} concerns itself only with the \textit{SOC} consumption, unlike the proposed strategy \eqref{equ:hyb_opt_prob} which also accounts for passenger comfort (characterized by the second and third terms in the cost function \eqref{equ:hyb_opt_prob_cost}) and avoids overly frequent gearshifts (through the constraint \eqref{equ:hyb_opt_prob_g5}). The reason for considering such a global \textit{SOC} consumption minimization solution is that it quantifies an upper limit on the energy efficiency performance, which will be used as a benchmark for evaluating performance of the other approaches. 

We use a dynamic programming (DP) algorithm \cite{sundstrom2009generic} to solve \eqref{equ:DP}, where the constraints \eqref{equ:hyb_opt_prob_c3}-\eqref{equ:hyb_opt_prob_c5} are handled through penalties.

Next, to highlight the benefit of employing multi-speed transmissions in BEVs for improving energy efficiency, we consider a BEV with a single reduction gear of ratio $i_{\text{g}}^{\text{s}}$ for forward driving. Its speed trajectory is optimized through repeatedly solving the following problem in a receding-horizon manner:
\begin{subequations}\label{equ:separated_opt}
\begin{align}
    \min \quad & J_{t}^{\text{sg}} = \sum_{k=0}^{N-1} \big( w_1'\,(v_{k+1|t}-v_{\text{r},k+1|t})^2 \nonumber \\[-1pt]
    &\quad\quad\quad  + w_2'\,(T_{\text{w},k|t}-T_{\text{w},k-1|t})^2 \big), \label{equ:separated_opt_c} \\[2pt]
    \text{s.t.} \quad &\quad  \eqref{equ:hyb_opt_prob_c1}-\eqref{equ:hyb_opt_prob_c5},
\end{align}
\end{subequations}
with respect to $T_{\text{w},k|t}$, $k = 0,\cdots,N-1$, where $T_{\text{w},k|t} = T_{\text{m},k|t}\, i_{\text{g}}^{\text{s}}$ instead of following \eqref{equ:wheel_torque}. The cost function \eqref{equ:separated_opt_c} is motivated by the observation that energy efficiency can be improved by smoothing speed profiles and reducing battery current spikes, which has been shown by the results in \cite{han2019fundamentals,li2016sequential,han2020acc}. 

The advantage of minimizing \eqref{equ:separated_opt_c} is that battery \textit{SOC} dynamics \eqref{equ:SoC} are not involved, so the number of states is reduced and estimating $V_{\text{oc}}$, $R_{\text{b}}$ and $\eta_{\text{m}}$ values through lookup tables $+$ interpolations is not needed, and thus, computations are simplified. It is shown through simulation case studies in Section~\ref{sec:5} that a speed trajectory determined based on \eqref{equ:separated_opt} can achieve, depending on driving cycles, $2.56\% \sim 10.42\%$ energy savings compared to following $v_{\text{r}}$ exactly. The results corresponding to \eqref{equ:separated_opt} and a single reduction gear are referred to as \textit{Optimized speed} \& \textit{Single gear}, and those corresponding to following $v_{\text{r}}$ exactly and a single gear are referred to as \textit{Baseline}.

\begin{figure} [ht!]
\begin{center}
 \includegraphics[width=80mm]{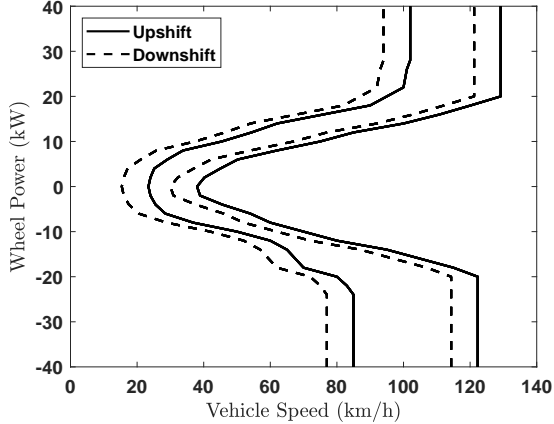}
 \caption{Gear shift map optimized through the approach of \cite{han2019optimized}.}
 \label{shift_map}
\end{center}
\end{figure}

Finally, for BEVs equipped with multi-speed transmissions, we show the benefit of optimizing vehicle speed and gearshift trajectories simultaneously using our proposed co-optimization strategy versus optimizing them separately. For the latter, we first optimize a static gear shift map offline following the approach of \cite{han2019optimized} (the obtained shift map is shown in Fig.~\ref{shift_map}), and then optimize the vehicle speed/wheel torque online based on the receding-horizon optimization \eqref{equ:separated_opt}. Specifically, after the pair of vehicle speed and desired wheel torque $(v_{t},T_{\text{w},t})$ has been determined through \eqref{equ:separated_opt}, the gear position $\eta_{\text{g},t}$ is selected according to the shift map and the motor torque is computed as $T_{\text{m},t} = \frac{T_{\text{w},t}}{i_{\text{g}}(\eta_{\text{g},t})\, i_0}$. The results corresponding to such a separate optimization procedure are referred to as \textit{Optimized speed} \& \textit{Shift map} (\textit{Map}).

\section{Results}\label{sec:5}

In this section, we evaluate performance of the proposed BEV speed and gearshift coordinated receding-horizon control strategy for improving energy efficiency through a comprehensive set of simulation case studies, and also compare it with the alternative energy efficiency optimization approaches described in Section~\ref{sec:4}.

Table~\ref{model_pram} summarizes the parameter values used for generating the results in this section. The values of parameters related to the vehicle and battery models \eqref{equ:system_model}-\eqref{equ:motor_speed}, as well as the maps for $V_{\text{oc}}(SOC)$, $R_{\text{b}}(SOC)$ and $\eta_{\text{m}} (T_{\text{m}},w_{\text{m}})$, are extracted from the high-fidelity powertrain simulation model \textit{ADVISOR} \cite{wipke1999advisor}. We use standard driving cycles to represent profiles of the reference speed $v_{\text{r}}$ for different driving conditions. The driving cycle is revealed gradually to the ego BEV as it drives forward, to model the real-time prediction of $v_{\text{r}}$ for a short time horizon.

\begin{table}[ht!]
\centering
\caption{Model Parameters.} \label{model_pram}
\begin{tabular}{|c|c|c|}
\hline
Symbol                              & Value [Unit]      \\ \hline
$m, m_{\text{eff}}$                                 & 1445 [kg]         \\ \hline
$r_\text{w}$                        & 0.3166 [m]        \\ \hline
$\rho$                              & 1.2 [$\text{kg/m}^{3}$]  \\ \hline
$A_\text{f}$                        & 2.06 [$\text{m}^{2}$]    \\ \hline
$C_\text{d}$                        & 0.312 [-]         \\ \hline
$g$                                 & 9.81 [$\text{m/s}^{2}$] \\ \hline
$\theta$                            & 0 [rad]           \\ \hline
$\mu$                               & 0.0086 [-]        \\ \hline
$C$                                 & 55 [Ah]           \\ \hline
$\eta_{\text{b}}^+, \eta_{\text{b}}^-$                & \{0.9, 1.11\} [-] \\ \hline
$\Delta t$                          & 1 [s] \\ \hline
$i_\text{g}(\eta_\text{g})$                       & \{3.05, 1.72, 0.92\}          \\ \hline
$i_0, i_\text{g}^\text{s}$                        & \{4.2, 7.2\} [-]                  \\ \hline
$\tau_{\text{min}}, \tau_{\text{max}}$    & \{1, 2\} [s] \\ \hline
$\delta_1, \delta_2, \varepsilon$      & \{5, 2, 0.1\} [-] \\ \hline
$\zeta_{\text{max}}$                      & 1 [-]             \\ \hline
$w_1, w_2, w_1', w_2'$         & \{$5 \times 10^{-4}, 2.5 \times 10^{-6}, 1, 10^{-3}$\}  \\ \hline
\end{tabular}
\vspace{0.5cm}
\end{table}

\begin{figure*} [ht!]
\begin{center}
  \includegraphics[width=125mm]{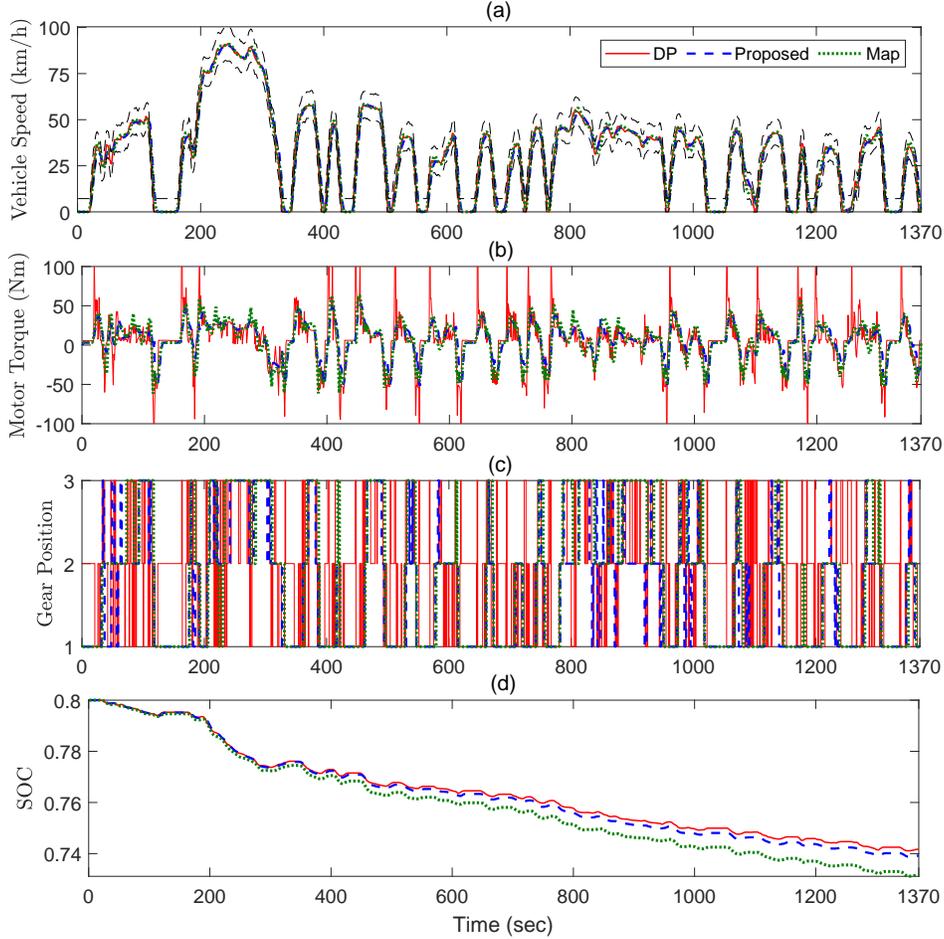}
  \caption{Simulation results for the UDDS driving cycle with planning horizon $N = 8$: (a) vehicle speed trajectories, (b) motor torque trajectories, (c) gear position trajectories, and (d) battery $SOC$ trajectories.}
  \label{UDDS_result}
\end{center}
\end{figure*}

Fig.~\ref{UDDS_result} illustrates the results for the Urban Dynamometer Driving Schedule (UDDS) cycle. Fig.~\ref{UDDS_result}(a) plots the obtained vehicle speed trajectories with different approaches, including the solution to \eqref{equ:DP} computed using DP, our proposed receding-horizon co-optimization solution, and the solution corresponding to separate optimization based on \eqref{equ:separated_opt} and the shift map in Fig.~\ref{shift_map}. Note that the speed constraints \eqref{equ:hyb_opt_prob_c4}, shown by the black dashed lines, are satisfied by all the approaches. Note also that the distance constraints \eqref{equ:hyb_opt_prob_c3} guarantee that the total travel distances corresponding to different approaches are close to each other (the difference is within $5$ meters). This ensures the \textit{SOC} consumption results of different approaches to be comparable. Fig.~\ref{UDDS_result}(b) -- (d) display, respectively, the corresponding motor torque, gear position, and battery \textit{SOC} trajectories.

It can be observed from Fig.~\ref{UDDS_result}(d) that the DP solution consumes the least \textit{SOC}, followed by our proposed solution (referred to as \textit{Speed-and-gearshift Co-optimization}), and the solution of \textit{Optimized speed} \& \textit{Shift map} consumes the most \textit{SOC}. Note that the DP solution relies on the assumption that $v_{\text{r}}$ over the entire trip is known a priori, which is hard to enforce in practice, while the other two approaches rely only on short-term predictions of $v_{\text{r}}$, which has been shown to be possible \cite{lefevre2014comparison,hyeon2019short,liu2019vehicle}. Moreover, the DP solution allows large wheel torque changes and arbitrarily frequent gearshifts in order to minimize \textit{SOC} consumption. As a result, we can observe significant spikes in the motor torque trajectory and a high frequency of gearshifts in the gear position trajectory of the DP solution compared to the other two approaches.

Our proposed solution consumes considerably less \textit{SOC} than the solution of \textit{Optimized speed} \& \textit{Shift map}, corresponding to $6.29\%$ improvement, and consumes only slightly more than the DP solution. This shows the effectiveness of our proposed receding-horizon control strategy for improving BEV energy efficiency and the superiority of speed and gearshift co-optimization over separate optimization. Specifically, for generating this result, we use $\zeta_{\max} = 1$ and $N = 8$, i.e., allow at most $1$ gear change over a planning horizon of $8\,$[s]. We choose $\zeta_{\max} = 1$ to balance the tradeoff between energy savings performance and computational complexity. For larger values of $\zeta_{\max}$, further improvements in energy efficiency are not significant. Note that allowing at most $1$ gear change over the planning horizon where the change can take place at any time instant of the horizon is more flexible than the assumption of constant gear position over the horizon in \cite{guo2016optimal}. The latter restricts the place over the planning horizon where the gear can be shifted to the beginning of the horizon.

In addition to the UDDS cycle, we also consider three other driving cycles and summarize the \textit{SOC} consumption results of all the approaches described in this paper for those driving cycles in Table~\ref{Tbl_results}. All simulations are conducted with the initial $SOC = 80\%$. 

\begin{table*}[ht!]
\centering
\caption{Battery $SOC$ consumption ($\%$) and improvement compared to \textit{Baseline} of different control approaches for different driving cycles.}
\label{Tbl_results}
\begin{tabular}{|c|c|c|c|c|c|c|c|c|c|c|c|c|c|}
\hline
\multirow{3}{*}{}                                                    & \multirow{3}{*}{Baseline} & \multicolumn{2}{c|}{\multirow{3}{*}{DP}}                                                        & \multirow{3}{*}{\begin{tabular}[c]{@{}c@{}} $N$ \end{tabular}} & \multirow{3}{*}{\begin{tabular}[c]{@{}c@{}}Optimized speed \\ \& Single gear\end{tabular}} & \multirow{3}{*}{\begin{tabular}[c]{@{}c@{}}Optimized speed \\ \& Shift map\end{tabular}} & \multicolumn{4}{c|}{Proposed}                                                                       \\ \cline{8-11} 
                                                                     &                           & \multicolumn{2}{c|}{}                                                                           &                                                                     &                                                                                            &                                                                                          & \multicolumn{2}{c|}{\multirow{2}{*}{$\Delta SOC$ (\%)}} & \multicolumn{2}{c|}{Computation time (sec)} \\ \cline{10-11} 
                                                                     &                           & \multicolumn{2}{c|}{}                                                                           &                                                                     &                                                                                            &                                                                                          & \multicolumn{2}{c|}{}                                 & $~~$average$~~$        &        worst        \\ \hline
\multirow{2}{*}{UDDS}                                                & \multirow{2}{*}{7.47}     & \multicolumn{2}{c|}{\multirow{2}{*}{\begin{tabular}[c]{@{}c@{}}5.83\\ (21.95\%)\end{tabular}}}  & 5                                                                   & 7.07 (5.35\%)                                                                              & 6.83 (8.57\%)                                                                            & \multicolumn{2}{c|}{6.37 (14.73\%)}                   &  0.73  &  1.39       \\ \cline{5-11} 
                                                                     &                           & \multicolumn{2}{c|}{}                                                                           & 8                                                                   & 6.78 (9.24\%)                                                                              & 6.55 (12.32\%)                                                                           & \multicolumn{2}{c|}{6.08 (18.61\%)}                   & 1.58  &  2.70       \\ \hline
\multirow{2}{*}{WLTC}                                                & \multirow{2}{*}{17.55}    & \multicolumn{2}{c|}{\multirow{2}{*}{\begin{tabular}[c]{@{}c@{}}15.0\\ (14.53\%)\end{tabular}}}  & 5                                                                   & 17.1 (2.56\%)                                                                              & 16.72 (4.73\%)                                                                           & \multicolumn{2}{c|}{16.17 (7.86\%)}                   &  0.63  &  1.14        \\ \cline{5-11} 
                                                                     &                           & \multicolumn{2}{c|}{}                                                                           & 8                                                                   & 16.3 (7.12\%)                                                                              & 16.01 (8.77\%)                                                                           & \multicolumn{2}{c|}{15.74 (10.31\%)}                  &  1.52  &  3.23         \\ \hline
\multirow{2}{*}{LA92}                                                & \multirow{2}{*}{12.86}    & \multicolumn{2}{c|}{\multirow{2}{*}{\begin{tabular}[c]{@{}c@{}}10.12\\ (21.31\%)\end{tabular}}} & 5                                                                   & 11.79 (8.32\%)                                                                             & 11.43 (11.12\%)                                                                          & \multicolumn{2}{c|}{10.72 (16.64\%)}                  &   0.64  &  1.49          \\ \cline{5-11} 
                                                                     &                           & \multicolumn{2}{c|}{}                                                                           & 8                                                                   & 11.52 (10.42\%)                                                                            & 11.17 (13.14\%)                                                                          & \multicolumn{2}{c|}{10.48 (18.51\%)}                  & 1.51  &  2.85           \\ \hline
\multirow{2}{*}{\begin{tabular}[c]{@{}c@{}}US06 \\ HWY\end{tabular}} & \multirow{2}{*}{9.43}     & \multicolumn{2}{c|}{\multirow{2}{*}{\begin{tabular}[c]{@{}c@{}}8.28\\ (12.2\%)\end{tabular}}}   & 5                                                                   & 9.01 (4.45\%)                                                                              & 8.88 (5.83\%)                                                                            & \multicolumn{2}{c|}{8.74 (7.32\%)}                    &   0.71  &  3.07                \\ \cline{5-11} 
                                                                     &                           & \multicolumn{2}{c|}{}                                                                           & 8                                                                   & 8.91 (5.51\%)                                                                              & 8.78 (6.89\%)                                                                            & \multicolumn{2}{c|}{8.68 (7.95\%)}                    & 1.85  &  3.17            \\ \hline
\end{tabular}
\end{table*}

As expected, the DP solutions consume the least \textit{SOC} for all of the four driving cycles. Our proposed \textit{Speed-and-gearshift Co-optimization} strategy is the second best, with, depending on the cycles, $7.32\% \sim 18.61\%$ improvements compared to \textit{Baseline} and $1.06\% \sim 6.29\%$ better than \textit{Optimized speed} \& \textit{Shift map}. In particular, except for the Worldwide Harmonized Light-duty Vehicles Test Cycles -- Class 3 (WLTC), \textit{Speed-and-gearshift Co-optimization} with a shorter planning horizon of $N = 5$ even outperforms the other two approaches, \textit{Optimized speed} \& \textit{Single gear} and \textit{Optimized speed} \& \textit{Shift map}, with a longer planning horizon of $N = 8$.

Among the four driving cycles, the benefit of employing multi-speed transmissions for improving energy efficiency is most significant for the UDDS cycle. This is because the UDDS cycle represents urban driving conditions, including many speed changes and several stop \& go maneuvers. For the UDDS cycle, our proposed \textit{Speed-and-gearshift Co-optimization} strategy achieves, respectively, $9.38\%$ and $6.16\%$ more energy savings when $N = 5$, and $9.37\%$ and $6.29\%$ more energy savings when $N = 8$, than the other two approaches. We also remark that because different approaches rely on different cost functions, for a fair comparison, when generating the results in Table~\ref{Tbl_results}, the weights for different terms of their cost functions have been tuned with trial and error to achieve their best energy efficiency performance.

\begin{figure} [ht!]
\centering
  \includegraphics[width=\columnwidth]{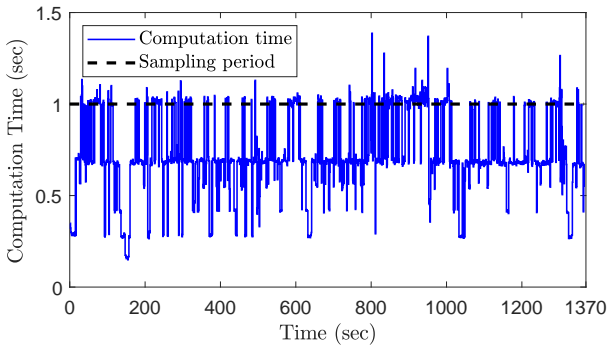}
  \caption{The computation time for obtaining our numerical solution at each time instant over the UDDS driving cycle.}
  \label{UDDS_Comp}
\end{figure}

To demonstrate the potential of our vehicle speed and transmission  gearshift coordinated receding-horizon control approach based on the proposed relaxation technique for real-time implementation, we plot the computation time for obtaining the numerical solution at each time instant over the UDDS cycle in Fig.~\ref{UDDS_Comp}, and summarize the average and worst computation times for the other cycles in Table~\ref{Tbl_results}. The computations are performed on the MATLAB R2018a platform running on an Intel Xeon E3-1246 3.50-GHz PC with 16.0-GB RAM. The NLP problems are solved using the MATLAB \textit{fmincon} function with the SQP method \cite{gill2012sequential}. The computation times are calculated using the MATLAB \textit{tic-toc} command. It can be seen that the computational cost is at a level that is comparable to the time available for real-time implementation. We remark that there are various ways to further reduce the computation times, for instance, by implementing the computations in $C$ \cite{altman2014accelerating}, replacing \textit{fmincon} with more efficient or tailored NLP solvers, exploiting inexact and real-time iteration solution strategies \cite{henriksson2004flexible,dontchev2018inexact} as well as symbolic and software optimization techniques \cite{walker2016design}. The investigation into these methods for further reducing the computational cost of our approach is left to future work.

\begin{figure} [ht!]
\centering
  \includegraphics[width=\columnwidth]{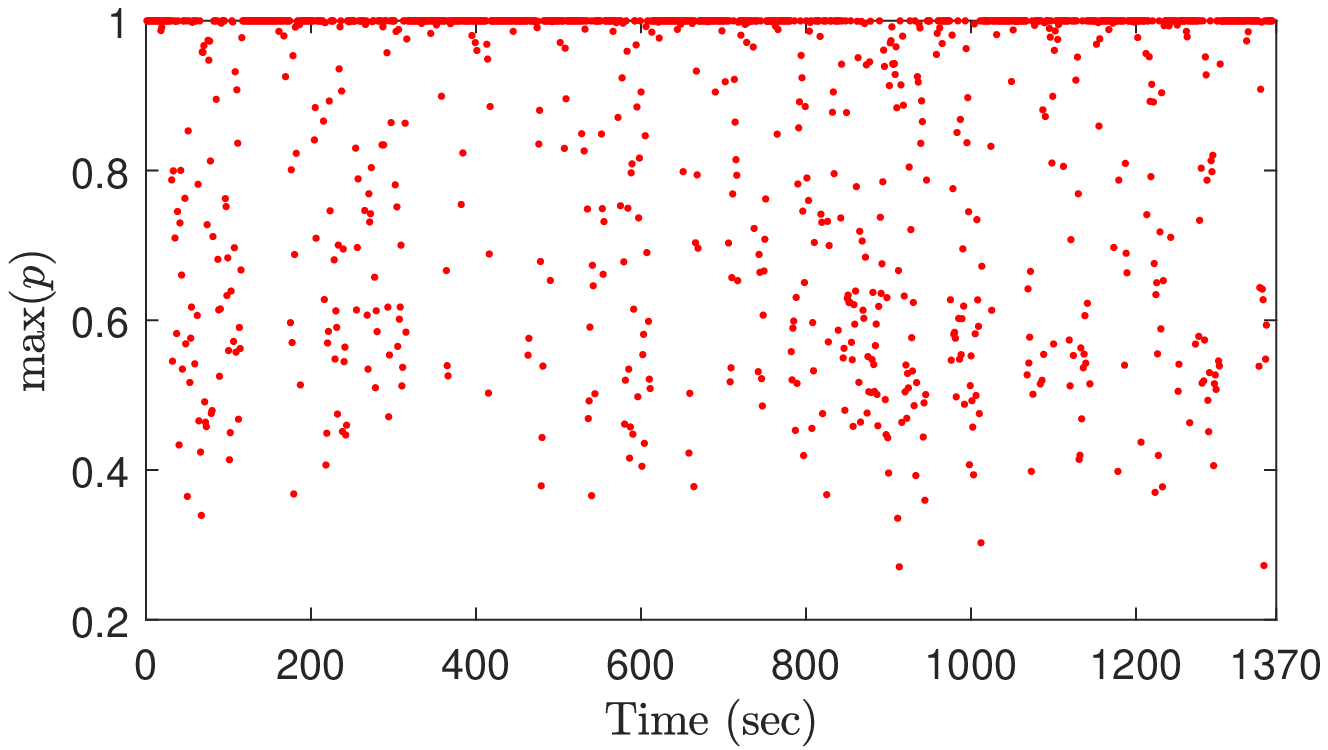}
  \includegraphics[width=\columnwidth]{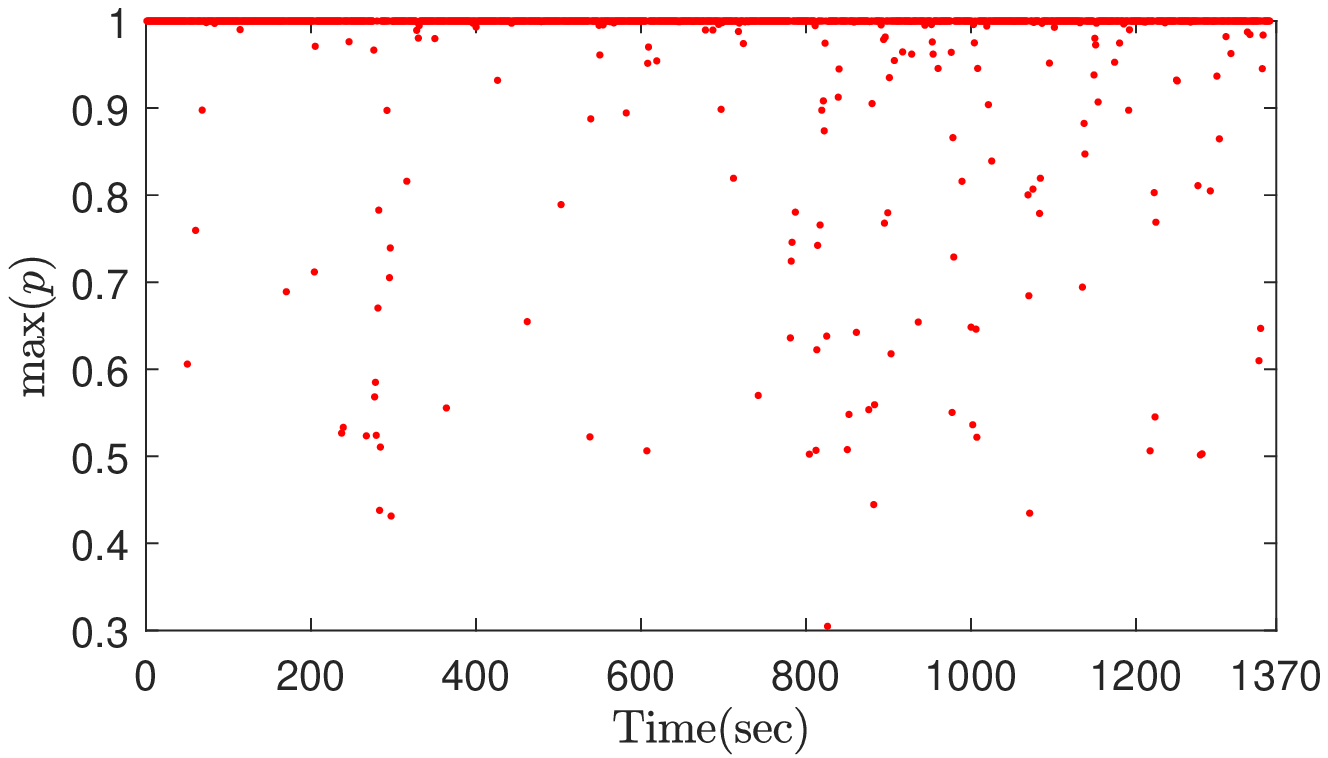}  
  \caption{The largest entry of the auxiliary vector $p$ of our numerical solution at each time instant over the UDDS driving cycle: (a) when the maximum number of SQP iterations is $50$; (b) when the maximum number of SQP iterations is $1000$.}
  \label{UDDS_Prob}
\end{figure}

Lastly, in Remark~2 we claimed that for the majority of time instants the numerical solutions to the relaxed version of our speed and gearshift co-optimization problem \eqref{equ:hyb_opt_prob_tf} should have integer $\bar{p}$'s. Proposition~3 ensures these solutions to be at least local minimizers of the original problem \eqref{equ:hyb_opt_prob_tf}. To confirm Remark~2, we plot the largest entry of the vector $\bar{p}$, $\max(\bar{p})$, of our numerical solution at each time instant over the UDDS cycle in Fig.~\ref{UDDS_Prob}. In particular, Fig.~\ref{UDDS_Prob}(a) shows the $\max(\bar{p})$ when the maximum number of SQP iterations is specified as $50$. This is the setting for generating the above \textit{SOC} consumption and computation time results. In this case, $62.8\%$ data points have $\max(\bar{p}) \approx 1$, where we categorize $\max(\bar{p}) \approx 1$ if $\max(\bar{p}) \in (0.95,1]$. Indeed, the deviations from $1$ of many data points are due to numerical errors. This is verified by Fig.~\ref{UDDS_Prob}(b), which shows the $\max(\bar{p})$ when the maximum number of SQP iterations is specified as $1000$. In this case, $92.8\%$ data points have $\max(\bar{p}) \in (0.95,1]$. Such an observation confirms our claim of Remark~2. We also remark that when the maximum number of SQP iterations is increased to $1000$, the computation time is also significantly increased, but with negligible \textit{SOC} consumption improvement. Therefore, $50$ is recognized as sufficient, and is also recommended, as the maximum number of SQP iterations for practical purpose. 

\section{Conclusions}\label{sec:6}

In this paper, we proposed a receding-horizon control strategy that simultaneously optimized vehicle speed and transmission gearshift trajectories for BEVs to achieve improved energy efficiency. 

The speed and gearshift co-optimization problem was formulated as an MINLP problem. To handle this MINLP problem, we proposed a novel continuous relaxation technique, transforming the original MINLP problem to a continuous optimization problem, so that approximate solutions to the original problem were obtained through solving the relaxed problem using off-the-shelf NLP solvers. Several theoretical results with respect to the feasibility and optimality correspondences between the original MINLP and the proposed relaxation have been discussed.

We applied the proposed continuous relaxation technique to solving the speed and gearshift receding-horizon co-optimization problem. Through a comprehensive set of simulation case studies and comparisons to several other energy efficiency optimization approaches, we showed that co-optimizing speed and gearshift could achieve considerably greater energy efficiency than optimizing them separately. We also showed that the proposed relaxation technique could reduce the online computational cost to a level that had the potential for real-time implementation. 

The proposed continuous relaxation technique may also be applied to other control-related problems involving systems with discrete modes, such as vehicle speed and transmission gearshift coordinated control for conventional internal combustion engine vehicles, vehicle speed and operation mode coordinated control for hybrid electric vehicles, etc. These are left as topics for future research.

\bibliographystyle{IEEEtran}
\bibliography{ref}

\end{document}